\documentclass[twocolumn,pra,superscriptaddress]{revtex4-2}

\usepackage[utf8]{inputenc}
\usepackage{float}
\usepackage{placeins}
\usepackage{qcircuit}
\usepackage{braket}
\usepackage{amsmath}
\usepackage[normalem]{ulem}
\usepackage{todonotes}
\usepackage{xcolor}
\usepackage{multirow}
\usepackage{braket}

\usepackage{bm}
\usepackage{bbold}

\usepackage{todonotes}

\newcommand{\ketbra}[2]{\mbox{$|#1\rangle\langle #2|$}}

\def\ketbra#1#2{{\vert#1\rangle\!\langle#2\vert}}

\def\braket#1#2{{\langle#1\vert#2\rangle}}

\DeclareMathOperator*{\argmin}{arg\,min}

\def\2LH{{\sc $2$-local Hamiltonian}}
\def\5LH{{\sc $5$-local Hamiltonian}}

\usepackage{dsfont}

\usepackage{mathtools} 
\graphicspath{{Figures/}}

\begin{document}

\title{Ion native variational ansatz for quantum approximate optimization}

\author{Daniil Rabinovich}\affiliation{Skolkovo Institute of Science and
Technology, Moscow, Russian Federation}
\author{Soumik Adhikary}\affiliation{Skolkovo Institute of Science and
Technology, Moscow, Russian Federation}
\author{Ernesto Campos}\affiliation{Skolkovo Institute of Science and
Technology, Moscow, Russian Federation}
\author{Vishwanathan Akshay}
\affiliation{Skolkovo Institute of Science and
Technology, Moscow, Russian Federation}
\author{Evgeny Anikin}\affiliation{The Russian Quantum Center, Moscow, Russian Federation}
\author{Richik Sengupta}\affiliation{Skolkovo Institute of Science and
Technology, Moscow, Russian Federation}
\author{Olga Lakhmanskaya}\affiliation{The Russian Quantum Center, Moscow, Russian Federation}
\author{Kiril Lakhmanskiy}
\affiliation{The Russian Quantum Center, Moscow, Russian Federation}
\author{Jacob Biamonte}\affiliation{Skolkovo Institute of Science and
Technology, Moscow, Russian Federation}
\homepage{http://quantum.skoltech.ru}

\thanks{Data and source code available upon reasonable request.}

\begin{abstract}
Variational quantum algorithms involve training parameterized quantum circuits using a classical co-processor. An important variational algorithm, designed for combinatorial optimization, is the quantum approximate optimization algorithm. Realization of this algorithm on any modern quantum processor requires either embedding a problem instance into a Hamiltonian or emulating the corresponding propagator by a gate sequence.  For a vast range of problem instances this is impossible due to current circuit depth and hardware limitations. Hence we adapt the variational approach---using ion native Hamiltonians---to
create ansatze families that can prepare the ground states of more general problem Hamiltonians. We analytically determine symmetry protected classes that make certain problem instances inaccessible unless this symmetry is broken.  We exhaustively search over six qubits and consider upto twenty circuit layers, demonstrating that symmetry can be broken to solve all problem instances of the Sherrington-Kirkpatrick Hamiltonian.  Going further, we numerically demonstrate training convergence and level-wise improvement for up to twenty qubits.  Specifically these  findings widen the class problem instances which might be solved by ion based quantum processors.  Generally these results serve as a test-bed for quantum approximate optimization approaches based on system native Hamiltonians and symmetry protection.  
\end{abstract}

 \maketitle


Modern quantum processors face noise and other imperfections that serve to limit realizable circuit depth.  Today's quantum processors allow realization of a short circuit with fixed error tolerance and tunable gates. This prompted many researchers to envision such quantum circuits as a new type of machine learning model \cite{Bia+17, KUB21, UKB20, MTB18} leading to variational quantum algorithms (VQA). In practice, open questions surround the expressability and trainability of low-depth circuits.  

A typical example of VQA, a so called Quantum Approximate Optimization Algorithm (QAOA), requires an alternating sequencing of two propagators to form approximate solutions of combinatorial optimization problems \cite{farhi2014quantum}. In such a setting, one of the propagators embeds the optimization problem as a diagonal unitary. With long enough circuits, QAOA can approximate the adiabatic evolution \cite{farhi2001quantum, Kadowaki1998,Boixo2014}, which certifies success. Moreover, QAOA has been shown to be successful in approximating solutions to several combinatorial problems even with medium depths \cite{farhi2014quantum,farhi2014quantumBOCP, wang2018quantum}. More optimistically, it was established by Lloyd \cite{lloyd2018quantum} and refined in~\cite{MBZ20} that QAOA anstaze circuits can be used to emulate any other quantum circuit. In practice, however, depth limits realizations of this approach in two interconnected ways.  Firstly, the QAOA sequence must be long enough to perform the minimization, thereby avoiding underparameterisation effects such as reachability deficits \cite{Aks+20,Aks+21}.  Secondly, the problem Hamiltonian itself can require significant overhead to realize, which in turn can translate into increasingly longer circuits.  

To take theory a step closer to practice, inspired in part by recent experiments \cite{Pagano2020}, we consider these challenges relative to the implementation of QAOA assuming an ion based quantum computer. Ion based quantum computers have made rapid progress in recent years \cite{Bruzewicz2019}, the primary advantages being (i) the ability to prepare identical qubits with all-to all-connectivity \cite{Cirac1995}, (ii) high gate fidelities \cite{Bermudez2017, Bruzewicz2019, Ringbauer2021, Pogorelov2021, Wright2019} and (iii) clean and relatively fast
readout \cite{Wang2021}. Ion based quantum computers enable a native realization of certain tunable Ising interactions \cite{Porras2004,Deng2005, Zhang2017, Maier2019, Monroe2021, Richerme2013, Richerme2013a, Richerme2014, Smith2016, Senko2014}. Such a Hamiltonian has recently been used in an experiment to implement a QAOA circuit with upto $40$ qubits and $2$ layers \cite{Pagano2020}. The problem Hamiltonian considered in this experiment is native to ion quantum computers. A natural question arises regarding the solution of more arbitrary combinatorial optimization problems.  Gate based realization of general problem Hamiltonians appears unattainable due to the required depths for all but the simplest cases.  How then might one solve general combinatorial optimisation problems using today's ion based processors? 

In an attempt to circumvent direct realisation \cite{Monroe2021, Sorensen1999, Pagano2020}, we are left to consider ion native Hamiltonians with tunable parameters. Whereas these native Hamiltonians would typically have no direct relation to the problem Hamiltonian, a variational ansatz can still be developed to solve a problem of interest. Following this idea we propose a native realization of variational QAOA-like circuit, and benchmark it against instances of Sherington-Kirkpatric (SK) Hamiltonian. 
This model arises in physics as a mean field approach to spin glasses -- a disordered system which exhibits non trivial magnetic behavior \cite{panchenko2012sherrington,PhysRevLett.35.1792}. Importantly, as the underlying interaction graph of SK Hamiltonian has all-to-all connectivity, it rules out a direct implementation for planar quantum processors. Furthermore, since the algorithmic performance of QAOA on typical instances of SK Hamiltonians can be estimated in the thermodynamic limit \cite{Farhi2019a}, the model therefore offers an ideal test case for the developments made here.  

For specific assignments of experimentally relevant parameters the proposed ansatz produces states with a type of reflection symmetry. 
This induces symmetry protection for some problem instances, that is these instances could not be minimized, as their ground state belongs to a different symmetry class. Violation of this symmetry with other assignments of the experimental parameters, however, allowed to minimize all SK problem instances. Thus, our approach extends the applicability of ion native QAOA to general combinatorial problems which have no native realization on ion based quantum processors.

\section{Quantum Approximate Optimization Algorithm on ion based quantum computers}

\subsection{ Traditional approach to QAOA }

The Quantum Approximate Optimization Algorithm approximates the ground state of a given $n$ qubit problem Hamiltonian $H_P$. That is, the algorithm prepares a unit norm quantum state $\ket{\psi}$ to find the minimum of the expectation value $\braket{\psi}{H_P|\psi}$. $H_P$ can embed a host of combinatorial problems for e.g.~graph minimization, Max-Cut \cite{farhi2014quantum} etc. The embedding is such that the solution to the problem is encoded in the ground state of $H_P$. 

The optimization begins with a candidate solution (a.k.a. QAOA ansatz):
\begin{align} \label{eq:QAOA_ansatz}
   &\ket{\Psi_p(\boldsymbol{\beta}, \boldsymbol{\gamma})} = \nonumber \\ 
   &\Big(\prod_{k=1}^p  \exp (- i \beta_k H_x) \exp (- i \gamma_k H_P) \Big) \ket{+}^{\otimes n},
\end{align}
where $H_x = \sum_{k=1}^n X_k$ is the mixer Hamiltonian, $X_k$ being the Pauli-$X$ operator on the $k$-th qubit. A $p$-depth QAOA circuit has $2p$ variational parameters $ \bm \beta, \bm \gamma \in [0, \pi)^{\times p}, [0,2\pi)^{\times p}$. The ground state of $H_P$ is then prepared by tuning $\bm \beta, \bm \gamma$ variationally following the minimization:
\begin{equation}\label{eq:opti}
     \min_{\boldsymbol{\beta}, \boldsymbol{\gamma}} \braket{\Psi_p(\boldsymbol{\beta}, \boldsymbol{\gamma})}{H_P | \Psi_p(\boldsymbol{\beta}, \boldsymbol{\gamma})},
\end{equation}
until a certain threshold is reached. A typical threshold, for example, is the spectral gap $\Delta$, of the considered problem Hamiltonian. Reaching this threshold guaranties certain overlap with the ground state \cite{akshay2022on} as, 

\begin{equation} \label{energy-ovelap}
    g(\psi) \geq 1 - \frac{\bra{\psi}H_P\ket{\psi} - \lambda_{min}}{\Delta}.
\end{equation}

\noindent Here, $ g(\psi)$ represents the ground state overlap with a state $\ket{\psi}$ and $\lambda_{min}$, the ground energy of $H_P$.

\subsection{Native gates and propagators in ion based quantum computers}

An ion based quantum computer comprises of ions in a radio frequency Paul trap \cite{Pogorelov2021, Schindler2013}. Certain electronic energy levels of the trapped ion approximate a two level system and hence serve as a qubit. Whereas there are multiple ways to control ion qubits, the most relevant realization is based on laser-ion interactions \cite{Bruzewicz2019}. We will therefore focus on this case for the rest of the paper.

Any unitary operation can be implemented on an ion based quantum computer using a universal set of native gates. The set comprises of two single qubit rotation operations $R_Z(\kappa)$ and $R_\phi (\theta)$ and one two qubit entangling gate---typically based on the M$\o{}$lmer-S$\o{}$rensen (MS) gate \cite{Sorensen1999,Sorensen2000}. 
Here $R_Z(\kappa) = \exp(-i \kappa Z)$, where $Z$ is the Pauli-$Z$ operator. The second single qubit gate is given by $R_\phi (\theta) = \exp(-i \theta S_\phi)$ where $S_\phi =  (X \cos\phi + Y \sin \phi)$, $X, Y$ being the Pauli-$X$ and Pauli-$Y$ operators. Both of these single qubit gates can be implemented by irradiating an ion with laser fields. The laser is characterized by specific parameters such as detuning, intensity, phase and laser pulse time, which can control the factors $\kappa$, $\theta$ and select the axis of rotation \cite{CrispinGardiner2015}.

The single qubit gates are known to demonstrate fidelities higher than $99.9 \%$ \cite{Akerman_2015, Ballance2016, Bermudez2017, Levine2018, Bruzewicz2019}. Two qubit gates on the other hand are more difficult to implement:  in general its performance degrades with the length of ion chains and is not uniform acrosss all qubit pairs. For ion chains of length 4 and 11, the two qubit gate fidelity are found to be ranging from 95$\%$ to 99$\%$ depending on the qubit pair 
\cite{Pogorelov2021, Wright2019}. 

Given the limited fidelities in the gate based model, implementing arbitrary circuits becomes a challenging task. One way to overcome this challenge is to realize native propagators directly instead of gates. In fact this approach is motivated by recent experimental demonstration of QAOA on 40 ions \cite{Pagano2020} and quantum simulation of the transverse-field Ising model with long-range interactions using 53 ion qubits \cite{Zhang2017}. It has been shown that upon the application of specific external laser fields a system of trapped ions can be described by the effective Ising Hamiltonian \cite{Monroe2021}:
\begin{equation}
     \label{eq:ising_hamiltonian}
  H_{I} = \dfrac{1}{2}\sum_{j\ne k} J_{jk} X_j X_k, 
\end{equation}
where the coupling constants $J_{jk}$ are approximated as
\begin{equation}
    \label{eq:ion_ising_couplings}
     J_{jk} \approx 
    \frac{J_\text{max} A_j A_k}{|j-k|^\alpha},
\end{equation}
with $A_j \in [-1,1]$ which can be controlled experimentally by varying the amplitude of the applied laser field. Here $J_\text{max} \in \mathbb{R}_+$ and $\alpha \in [0,3]$ \cite{Porras2004, Monroe2021}. $H_I$ and hence the entangling propagator $\exp(-i \gamma H_I)$ are native to the ion hardware. Implementing this propagator suppresses phonon excitation, which is one of the main source of heating and decoherence in ion quantum computers \cite{Zhang2017}. This makes the choice of $H_I$ pertinent. More experimentally relevant details on the ion Hamiltonian are provided in appendix~\ref{appen:ion_hamil}.

\subsection{Native QAOA implementation}

In the gate based model, implementation of the circuit \eqref{eq:QAOA_ansatz} on an ion based quantum computer would require the circuit to be converted into a sequence of native gates \cite{Schindler2013}. The propagator $\exp(- i \beta_k H_x)$ admits a decomposition in terms of single qubit rotation operation as $(R_{\phi=0} (\beta_k))^{\otimes n}$. Decomposition of the propagator $\exp (- i \gamma_k H_P)$ on the other hand can prove more challenging as it involves both single and two qubit gates. As discussed previously two qubit gates are significantly noisier than single qubit gates. Thus, only a limited number of two qubit gates can be implemented in order to restrict the total error. This makes realization of the propagator $\exp (- i \gamma_k H_P)$ for an arbitrary $H_P$, impossible, within a reasonable error tolerance. 



These shortcomings may however be avoided by choosing an alternate candidate ansatz to minimize $H_P$:
\begin{align} \label{eq:QAOA_ansatz_alt}
   &\ket{\psi_p(\boldsymbol{\beta}, \boldsymbol{\gamma})} = \nonumber \\
   &\prod_{k=1}^p \Big( \exp (- i \beta_k H_x) \exp (- i \gamma_k H) \Big) \ket{+}^{\otimes n},
\end{align}
where $H$ is a non-local Hamiltonian that can create entanglement and does not commute with $H_x$. Such an ansatz is reminiscent of Quantum Alternating Operator Ansatz, also QAOA \cite{Hadfield2019from}. 
Apart from these algorithmic requirements, the choice of $H$ should be dictated by specifications of the ion quantum computer such that the propagator $\exp(-i\gamma_k H)$  and hence the circuit \eqref{eq:QAOA_ansatz_alt} can be implemented naturally on the ion hardware. This step in particular allows us to circumvent the gate model while implementing QAOA. 
Consider the ansatz:
\begin{align}
\label{eq:QAOA_ansatz_exp}
    &\ket{\phi_p(\boldsymbol{\beta}, \boldsymbol{\gamma})} =\nonumber\\ 
    &\text{H}_+ \Big[\prod_{k=1}^p \Big( \exp(-i  \beta_k H_z) \exp(- i  \gamma_k H_I) \Big) \ket{0}^{\otimes n} \Big],
\end{align}
where $\text{H}_+ = (\ketbra{+}{0} + \ketbra{-}{1})^{\otimes n}$ is the Hadamard gate applied to each of the qubits and $H_z = \sum_k Z_k$. It can be shown that $\ket{\phi_p(\boldsymbol{\beta}, \boldsymbol{\gamma})}$ is equal to $\ket{\psi_p(\boldsymbol{\beta}, \boldsymbol{\gamma})}$ in \eqref{eq:QAOA_ansatz_alt} with 
\begin{align}
    H = \text{H}_+ H_I \text{H}_+ = \dfrac{1}{2}\sum_{j\ne k} J_{jk} Z_j Z_k
    \label{eq:H_ZZ}
\end{align} (see Appendix \ref{equality} for details). Eq.~\eqref{eq:QAOA_ansatz_exp} is therefore indeed a QAOA circuit. It gives rise to a variational state space 
\begin{equation}
    \mathcal{S} = \bigcup_{\bm\beta,\bm\gamma}\ket{\phi_p(\bm\beta, \bm\gamma)}.
\end{equation}
We further show that \eqref{eq:QAOA_ansatz_exp} is ion compatible; each term involved in the preparation of $\ket{\phi_p(\boldsymbol{\beta}, \boldsymbol{\gamma})}$ is native to the ion quantum computer. Using the propagator $\exp(- i  \gamma_k H_I)$ in \eqref{eq:QAOA_ansatz_exp} is vital; the propagator naturally arises when the system evolves under the ion Hamiltonian $H_I$, and thus requires no gate decomposition unlike the propagator $\exp(- i  \gamma_k H_P)$ in the original QAOA setting \eqref{eq:QAOA_ansatz}.
The term $\exp(-i  \beta_k H_z)$ can be decomposed into single qubit rotation operations which, as was discussed, can be implemented with high fidelity. Equivalently this term can also be implemented as a global rotation operation by applying a single laser on all the ions at once. We further note that the mixer Hamiltonian $H_z$ does not commute with $H_I$ as per the requirement of the algorithm. Finally the gate $\text{H}_+$ is implemented as $( R_{\phi ={\frac{\pi}{2}}}(\frac{\pi}{4}) R_Z(\frac{\pi}{2}))^{\otimes n}$, where all gates involved in the decomposition are again native to the ion based quantum computers.


\section{Numerical Experiments}
\subsection{The problem}

We benchmark the proposed algorithm with all instances of $n=6$-qubit Sherrington-Kirkpatrick (SK) Hamiltonians
\begin{equation}
    H_{P} = \frac{1}{2} \sum_{j\ne k} K_{jk} Z_j Z_k,
\end{equation}
where $K_{jk} \in \{ 1, -1 \}$. Our goal is to minimize $H_P$ with respect to the ansatz \eqref{eq:QAOA_ansatz_exp} (see Appendix D numerics details).
In other words, we seek for the elements of
\begin{equation}
    \argmin\limits_{\ket{\psi}\in\mathcal{S}}\bra{\psi}H_P\ket{\psi}.
\end{equation}
We consider the minimization to be successful if the difference between the energy $\bra{\psi}H_P\ket{\psi}$ and the true ground state energy of $\lambda_{min}(H_P)$ is upper bounded by $\dfrac{5}{100}|\lambda_{min}(H_P)|$, that is a $5 \%$ threshold on the ground state energy, which is always below the gap for any $6$-qubit SK instance.


During the minimization, we fix certain parameters in $H_I$. We work in the dimensionless units where energy is measured in $\mathrm{kHz}$ and propagation time in 
$\mathrm{ms}$ and set $J_{max}=4$ (see Appendix \ref{appen:ion_hamil} for justification). We chose $\alpha = 1$, which we show to be sufficient in the next section. 
Thus, we consider the couplings of the form 
\begin{equation}     
    J_{jk} = 
    \frac{4 A_j A_k}{|j-k|}.
    \label{eq:couplings_simp}
\end{equation}

\subsection{Results}

\subsubsection{Symmetric configuration}
\label{sec:allone}

We start with elementary experimental setting $A_j=1$ for each qubit $j$. The ion Hamiltonian couplings \eqref{eq:couplings_simp} under this assignment of $A_j$ become $J_{jk} = \frac{4}{|j-k|}$. Based on our numerical data, we observed that not all SK Hamiltonians could be minimized by our algorithm with the present setting. 

\begin{figure}[htp]
\centering
\includegraphics[width=\linewidth]{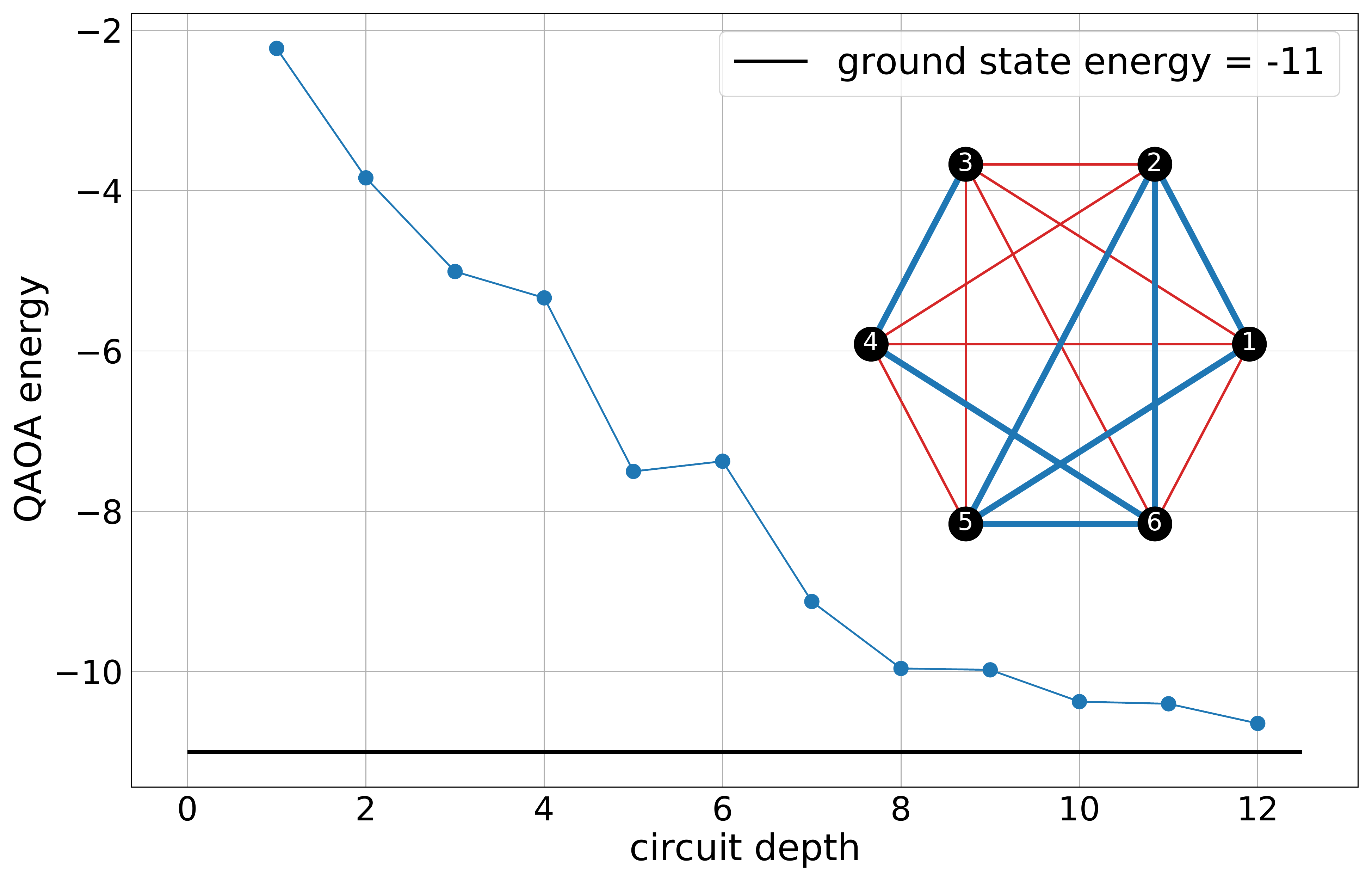}
\caption{(color online) Energy minimization of an easy instance with respect to the QAOA ansatz \eqref{eq:QAOA_ansatz_exp}. Here we set $A_j = 1$ for all qubits $j$. The energy (see \eqref{eq:opti}) gradually decreases with increasing depth until minimization. Insert: SK instance that is being minimized, red thin edges depict $K_{jk}=+1$, while blue thick ones depict $K_{jk} = -1$.}
\label{fig:good}
\end{figure}

\begin{figure}[htp]
\centering
\includegraphics[width=\linewidth]{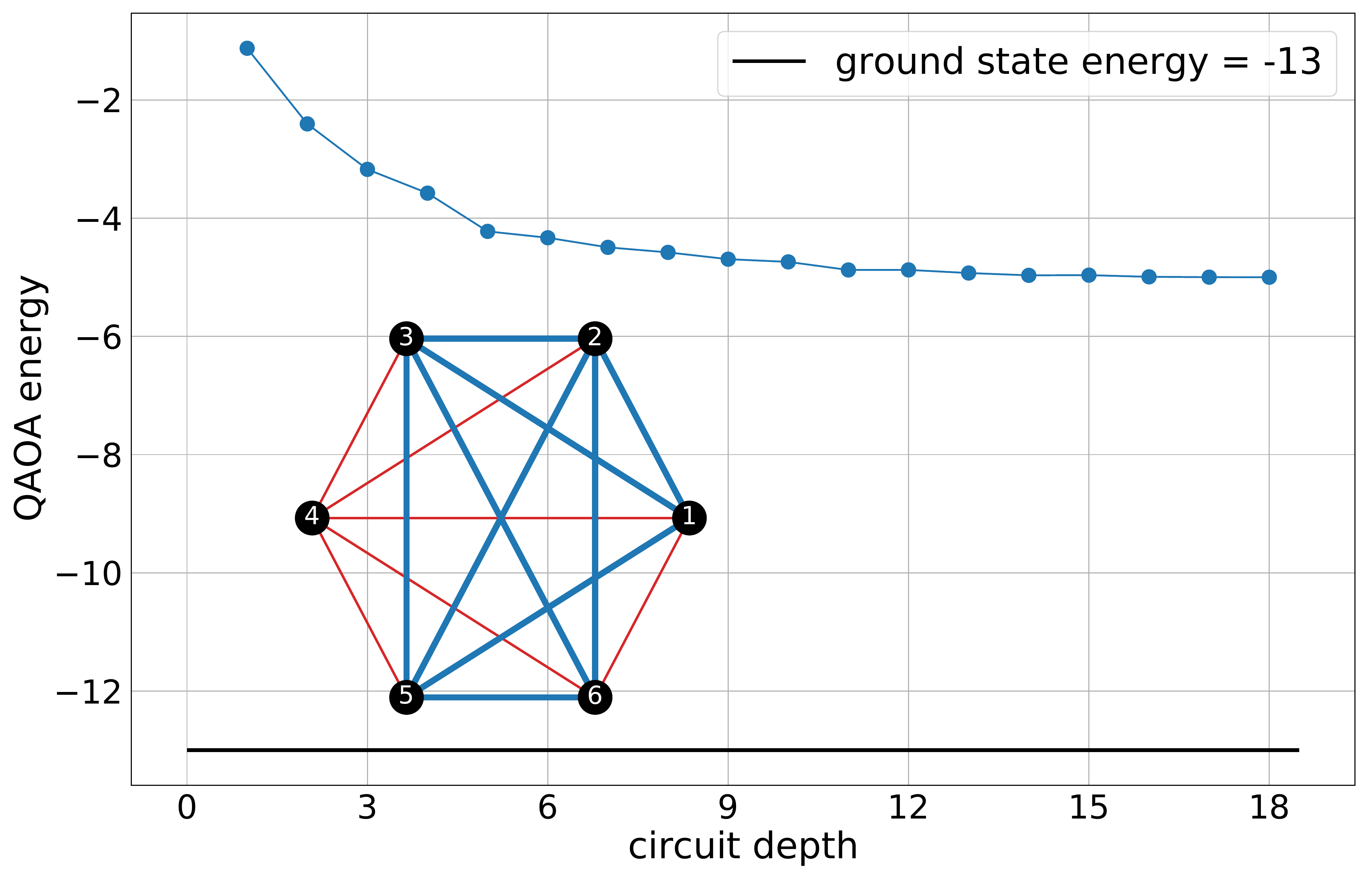}
\caption{(color online) Energy minimization of a hard instance with respect to the QAOA ansatz \eqref{eq:QAOA_ansatz_exp}.  Here we set $A_j = 1$ for all qubits $j$. The energy (see \eqref{eq:opti}) gradually decreases with increasing depth yet stagnates at some intermediate value (here $\sim -5$). Insert: SK instance that is being minimized, red thin edges depict $K_{jk}=+1$, while blue thick ones depict $K_{jk} = -1$.}
\label{fig:bad}
\end{figure}

The instances for which the minimization was successful will be called  {\it easy instances}. Figure \ref{fig:good} demonstrates the energy minimization for such a case: algorithmic performance gradually improves with increasing depth until the ground state is recovered. In contrary, figure \ref{fig:bad} illustrates the energy minimization of what we call {\it hard instances}. These instances could not be minimized by our algorithm in the symmetric setting: while performance still can improve with increasing depth, the algorithm stagnates and energy can not go past a certain threshold.

The reason why some of the SK instances could not be minimized is related directly to the symmetries intrinsic to our ansatz:
\begin{align}
    X^{\otimes n} \ket{\psi} = \ket{\psi}
    \label{eq:Z2}\\
    \mathcal{R} \ket{\psi} = \ket{\psi}
    \label{eq:R}
\end{align}
Here $\mathcal{R}$ is a linear reflection operation that acts on a bitstring
as $\mathcal{R} \ket{j_1 j_2 \cdots j_{n-1} j_n} = \ket{j_n j_{n-1} \cdots j_2 j_1}$. Physically, operation $\mathcal{R}$ corresponds to reflection of the ion string around the middle point. While the symmetry \eqref{eq:Z2} is intrinsic to our ansatz \eqref{eq:QAOA_ansatz_exp}, the symmetry \eqref{eq:R} arises only in the symmetric configuration when $A_j = A_{n-j+1}$. 



These relations \eqref{eq:Z2} and \eqref{eq:R} restrict the class of states that can be prepared by our QAOA ansatz. In-fact $\mathcal{S} \subseteq  \mathcal{V}$, where $\mathcal{V}$ is the space of all possible $n$ qubit states that satisfy \eqref{eq:Z2} and \eqref{eq:R} simultaneously. We call $\mathcal{V}$ the symmetric state space. Naturally, this induces a symmetry protection for the instances without symmetric ground state $\ket{\chi} \in \mathcal{V}$, all of which were therefore classified as hard. 
Moreover, we observe that for the case of $6$ qubits SK instances considered here, all instances with symmetric ground states are classified as easy. We further show in Appendix \ref{appen:sym_analysis} that one can define a basis in $\mathcal{V}$ which forms a minimal set of states, sufficient to minimize all easy instances.
This results demonstrates that the choice of $J_{max}$ and $\alpha$ was sufficient: with these choices the algorithm minimized all the instances with symmetric ground states. In contrary, the other instances could not be minimized for any other choice of $J_{max}$ and $\alpha$ due to symmetry protection. 

Finally, we observed that $\sim 0.67$ of all SK instances are hard due to symmetry protection induced by \eqref{eq:R}. This can be seen in figure \ref{fig:fraction} (orange curve): the fraction of solved instances monotonically grows with increasing depth until it stagnates at depth $p=12$ at the value $\sim 0.33$.
A question follows naturally: is it possible to have a setting where all SK instances could be minimized? We address this next.


\subsubsection{Asymmetric configurations}
We consider configurations that violate the condition $A_j = A_{n-j+1}$ and, therefore remove the symmetry restriction \eqref{eq:R}. We consider two such settings (i) randomly assigned values to the parameters $A_j$ and (ii) $A_j = 1$, $\forall j\ne j'$ while the value of $A_{j'}$ is selected at random.

\begin{figure}[htp]
\centering
\includegraphics[width=\linewidth]{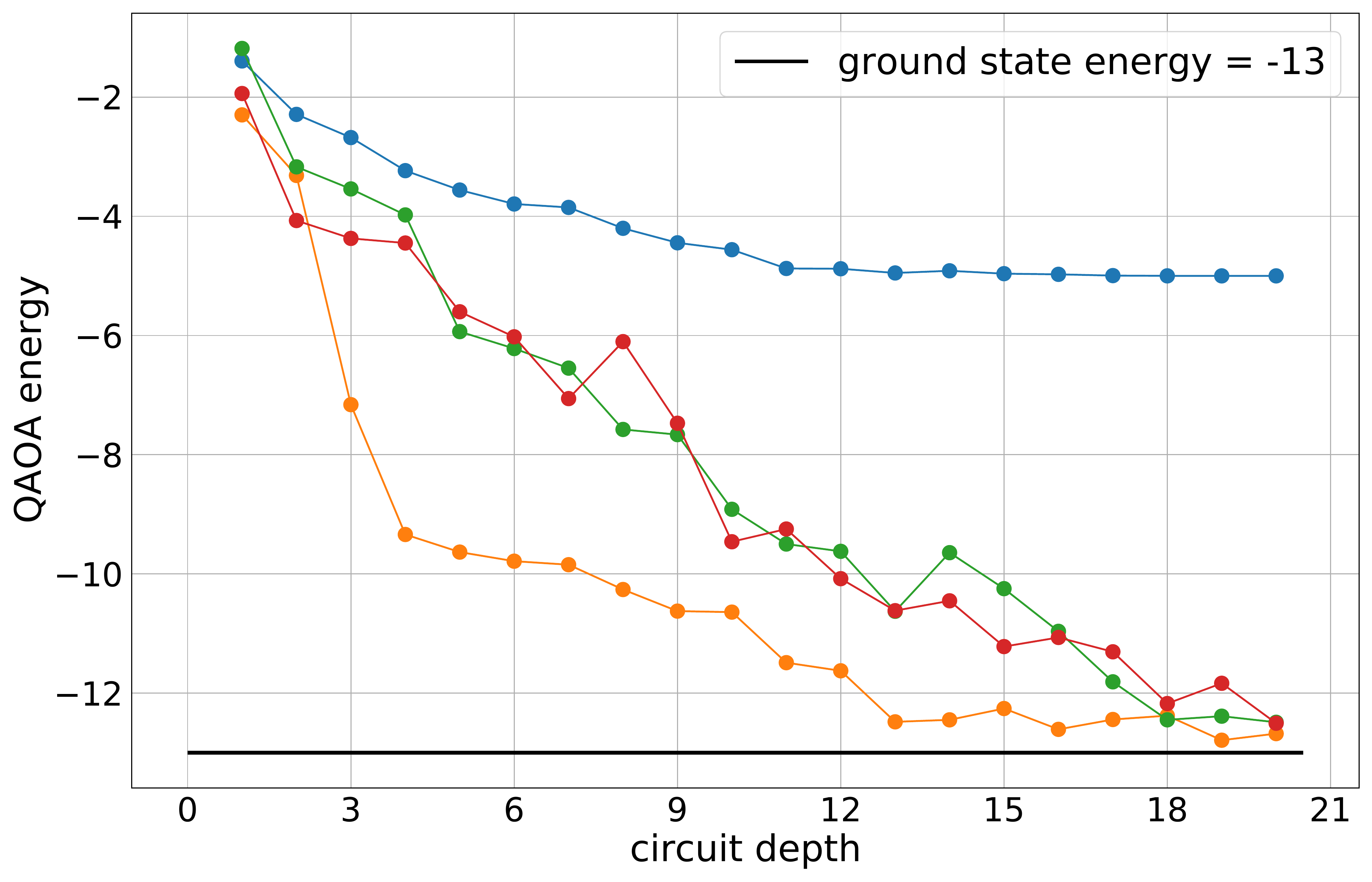}
\caption{(color online) Energy minimization for the instance considered in figure \ref{fig:bad}, with respect to the QAOA ansatz \eqref{eq:QAOA_ansatz_exp}. Here we demonstrate three random assignments of $A_j$, shown in different colors. We see that the energy monotonically decreases with depth and eventually reaches the minimum. We contrast this with the performance that was recorded with the symmetric configuration: $A_j=1$, depicted by the blue curve.}
\label{fig:bad_rand_omega}
\end{figure}

In both settings we observe that all SK instances are solved by our algorithm. The blue curve in figure \ref{fig:fraction} demonstrates an example of this, where the fraction of easy instances can be seen to be monotonically increasing with QAOA circuit depth, until it reaches 1 at depth $p=20$.
From this data we confirm that lifting the symmetry is sufficient to minimize all the instances including those that were classified as hard in the symmetric configuration. Minimization of such instances became possible as violation of \eqref{eq:R} ensures that $\mathcal{S}$ is no longer restricted to $\mathcal{V}$. Figure \ref{fig:bad_rand_omega} shows the QAOA recovered energy minimization for one such instance using setting (i). It must be noted here that despite similar performances setting (ii) is, in principle, easier to implement experimentally as fewer ions require individual addressing.


\subsubsection{Problem specific configuration}

\begin{figure}[htp]
\centering
\includegraphics[width=\linewidth]{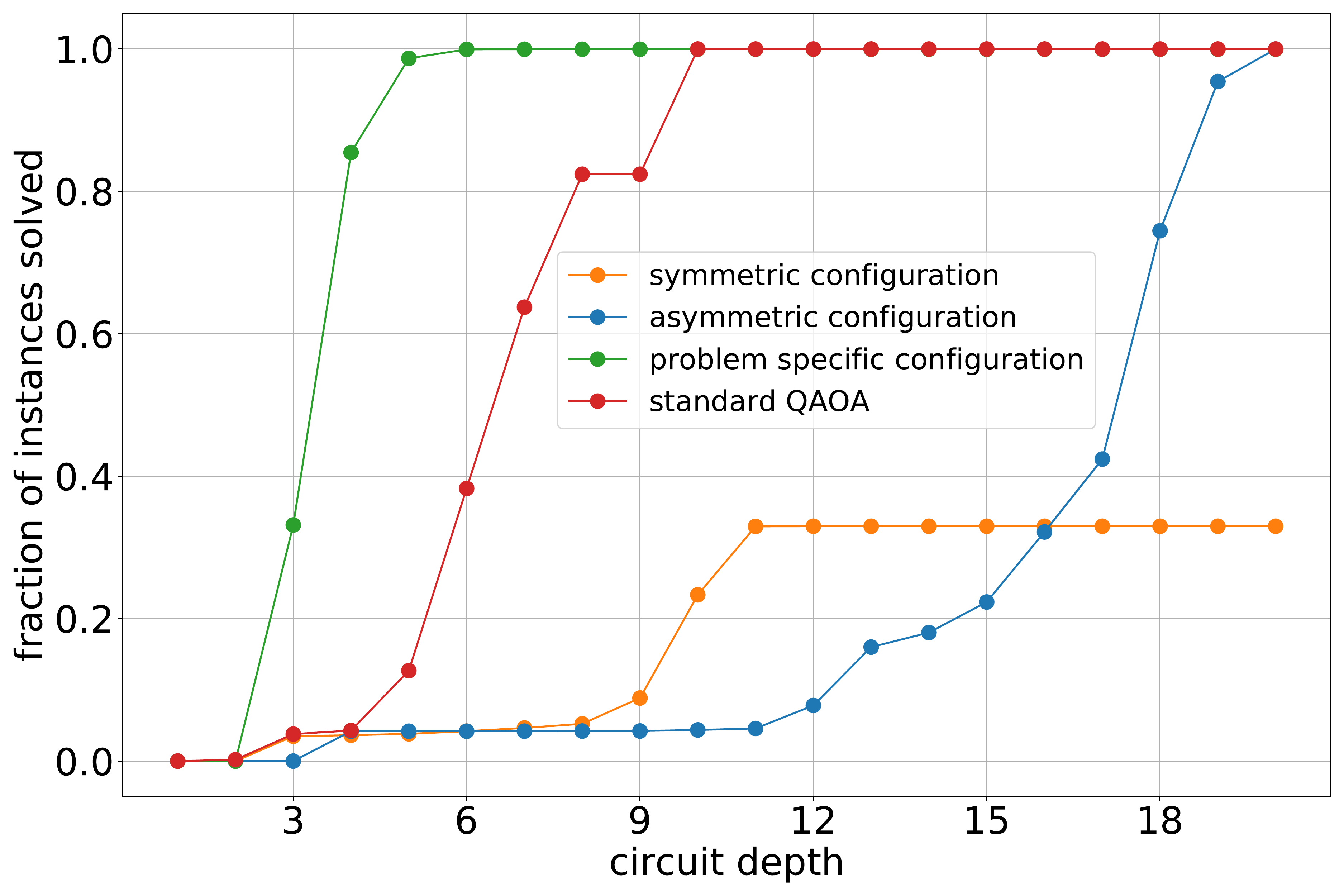}
\caption{(color online) Fraction of SK instances with $n=6$ that could be minimized by the proposed QAOA ansatz at each depth. For the orange curve we have set $A_j=1$ for all qubits $j$. For the blue curve we use randomly assigned values of $A_j$ provided $A_j \ne A_{n-j+1}$. For the green curve an instance was classified as solved if it was minimized with at least one of the 50 randomly sampled sets of $A_j$. The red curve illustrates fraction of SK instances that are minimized by standard QAOA circuit \eqref{eq:QAOA_ansatz}.}
\label{fig:fraction}
\end{figure}

We find that using problem specific assignment of potentially unequal $A_j$ can lead to faster minimization of the problem instances compared to the symmetric and asymmetric assignments. This is confirmed by our numerical data: for each instance there exists a specific assignment of $A_j$ which allows to minimize the instance in no more than $7$ layers (see figure~\ref{fig:fraction} green curve). Thus, we see a considerable reduction in the number of layers required to minimize any instance compared to (i) the previous settings, which require depths upto $p=20$, and (ii) standard QAOA which required depth $p=10$ to minimize all SK instances (see red curve in figure~\ref{fig:fraction}).

\subsubsection{Large system size}

\begin{figure}[htp]
\centering
\includegraphics[width=\linewidth]{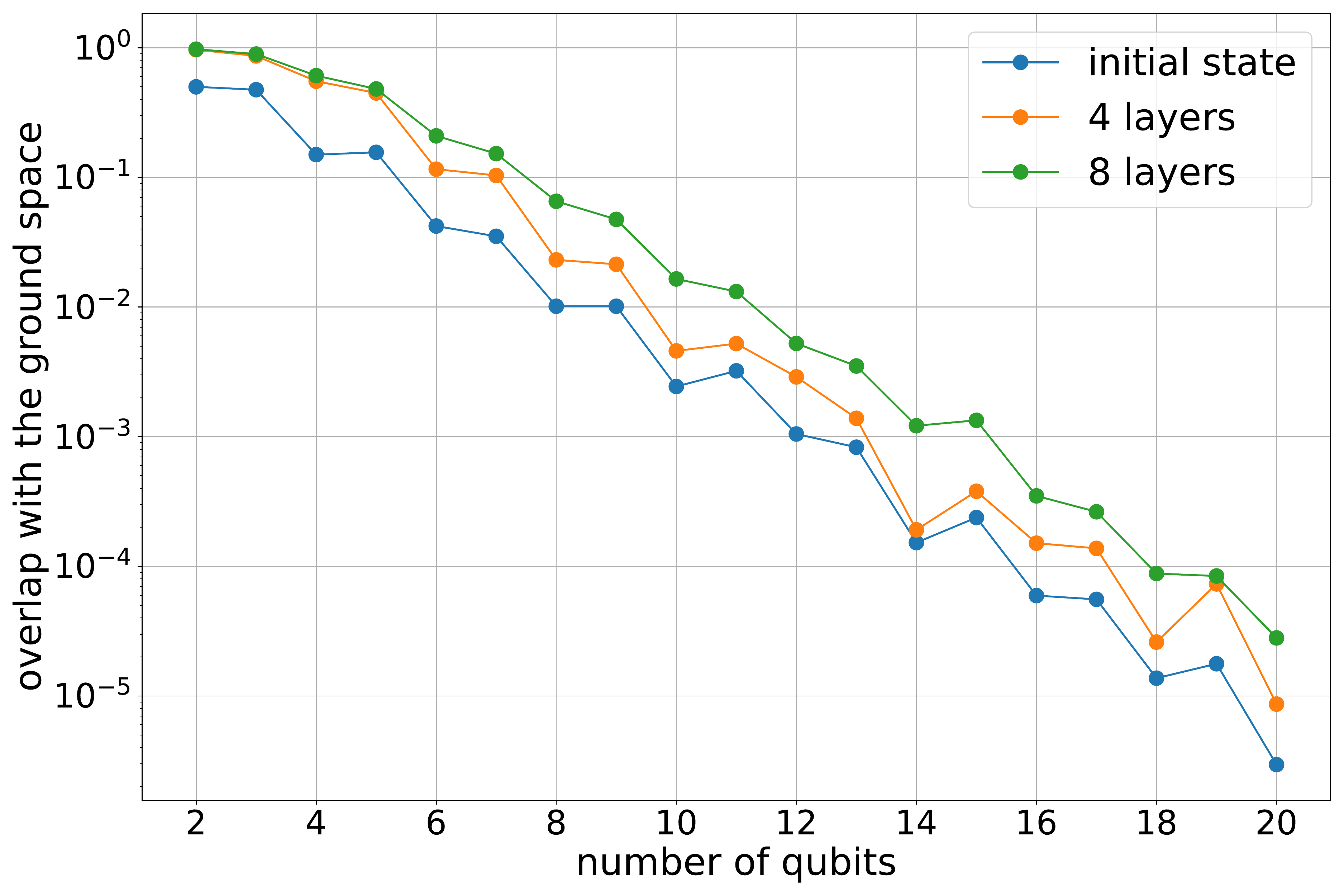}
\caption{(colors online) 
Overlap of the state prepared by the ansatz \eqref{eq:QAOA_ansatz_exp} with the ground space, for different number of qubits. At each $n$ the overlap is averaged over a set of fixed $20$ random instances. Blue curve represents the overlap calculated for the initial QAOA state $\ket{+}^{\otimes n}$. Orange curve illustrates overlap for the trained $p=4$ layer circuit. Green curve demonstrates improved overlap for $p=8$ layer circuit, obtained via repeated sampling of circuit parameters.}
\label{fig:20qubit}
\end{figure}

While the benchmarking of the proposed algorithm was set against $n=6$-qubits SK instances, here we demonstrate its performance for a larger system sizes. We consider minimization of random instances of the SK Hamiltonian with the number of qubits from $n=2$ to $20$ using  $p=4, 8$ layer QAOA-like circuit. For every considered instance a specific choice of $A_j$ was used. In this setting we demonstrate that the overlap of the prepared state with the ground space of the instances improves with increasing depth (see figure \ref{fig:20qubit}). This shows that the present algorithm has the potential to minimize arbitrary instances of SK Hamiltoninas for large problem sizes.

We note from figure \ref{fig:20qubit} that the overlap diminishes with the number of qubits and the scaling is approximately exponential. This limitation is not peculiar to our modification alone and is known in traditional QAOA \cite{Pagano2020}. This arises from the fact that QAOA, in general, might require exponential (in $n$) depth to minimize an $n$ qubit Hamiltonian \cite{2002quant.ph..6003V,2010PNAS..10712446A, 2017PhRvA..95f2317J}.


\section{Conclusion} 


We circumvent the traditional embedding problem and thereby expand the application potential for modern day ion based processors. Given a problem instance, one must either embed its Hamiltonian or emulate its propagator on a given quantum processor. This requires overheads that currently prove the task impossible in all but the simplest of cases. Hence we adapt the variational approach to consider an ansatze family of states, based on the native ions Hamiltonian, that prepare low energy states of the corresponding problem Hamiltonian.  



We exhaustively benchmarked the proposed algorithm by considering instances of Sherrington-Kirkpatrick Hamiltonians for $n=6$ qubits. Upon considering multiple experimentally relevant setups, we observed that certain configurations induce symmetry protection for specific classes of problem instances, which thus can not be minimized by the algorithm.
Configurations which violate the symmetry allowed the minimization of arbitrary SK instances. In real settings, the required asymmetry can equivalently come from the more general ion-compatible Hamiltonian used as a propagator, or from noise in the system.
We demonstrate that with problem specific configuration any instance can be minimized with no more then $7$ layers, which is experimentally accessible. Numerically we considered up to $n=20$ qubits and demonstrated both training and performance improvement with increasing circuit depth. 

Considering more general combinatorial optimization problem, one would expect the problem density (the constraint to variable ratio) induce under-parameterization which might necessitate large circuit depth \cite{Aks+20}. As training such circuits becomes costly at large depth,  certain heuristics based on the parameter concentrations \cite{akshay2021parameter} and layerwise training might be useful to simplify the optimization. While, in principle, the proposed algorithm can still suffer from training saturation, the performance might be recovered under certain types of noise \cite{campos2021training}. An alternative way to improve circuit expressivity is to variationally control the $A_j$ terms, which is a subject of further studies.

\begin{acknowledgements}
The work was supported in the framework of the Roadmap for Quantum computing (Contract No.~868-1.3-15/15-2021 and~R2163).  The large calculations were performed on the Zhores supercomputer. 
The code is written in Python with Intel optimised libraries and is available on reasonable request.
\end{acknowledgements}

\bibliographystyle{unsrt}
\bibliography{references}

\appendix

\section{The ion model Hamiltonian}
\label{appen:ion_hamil}
The review on different types of spin-spin interactions existing in trapped ion Coulomb crystals exposed to external laser fields can be found 
in \cite{Monroe2021}. Here we focus on a specific type of Ising couplings which arises when the ions are illuminated by bichromatic laser beams with same symmetric detunings $\Delta$ (from the qubit transition frequencies), and ion--dependent beams intensities. In the Lamb--Dicke approximation and in the dispersive regime \cite{Kim2009} the qubit dynamics can be described by the effective Hamiltonian \eqref{eq:ising_hamiltonian}
with the couplings
\begin{equation}
  \label{eq:ising_couplings_general}
  \begin{gathered}
    J_{jk} = \Omega_{j}\Omega_k \mathcal{C}_{jk}.
  \end{gathered}
\end{equation}

Here $\Omega_{j}$ are the Rabi frequencies induced by the laser field acting on ion number $j$ and are, in general, controlled 
individually for each ion. The matrix $\mathcal{C}_{jk}$ does not depend on $\Omega_j$ and has complex dependence on the phonon normal modes and laser detunings for each phonon mode frequency \cite{Monroe2021}. 

In this paper, we focus on the long range Ising couplings 
given by \eqref{eq:ising_hamiltonian} and  \eqref{eq:ion_ising_couplings}.
When the detuning $\Delta$ exceeds the frequency of the radial center of mass mode, 
the dependence of $\mathcal{C}_{jk}$ on $j$ and $k$ can be approximated with the power law $\mathcal{C}_{jk} \propto 1/|j-k|^\alpha$. 
Given the notation $A_{j} = \Omega_{j}/|\Omega_\mathrm{max}|$, and $J_\text{max} = J_{j,j+1}$ when $\Omega_j = \Omega_\mathrm{max}$ for all qubits $j$,
one gets \eqref{eq:ion_ising_couplings} from \eqref{eq:ising_couplings_general}.
The exponent of the power law $\alpha$ ranges between $0$ and $3$ depending on the value of $\Delta$. Nevertheless, the most relevant values for experimental realization lie between 0.8 and 1.8 \cite{Zhang2017}, which allow to avoid motional decoherence and experimental drifts.

Typically, $J_\text{max}\sim 1 \text{kHz}$ implying propagation times with the ion Hamiltonian are of the order of several milliseconds.  Thus, the standard range of QAOA angle $\gamma\in[0,2\pi)$ can be treated as propagation time measured in milliseconds. In dimensionless units we chose $J_{max}=4$ such that the phase $\gamma J_{max}$ could range in the experimentally accessible range of several multiples of $\pi$.

\section{Relation between $\ket{\phi_p(\boldsymbol{\beta}, \boldsymbol{\gamma})}$ and $\ket{\psi_p(\boldsymbol{\beta}, \boldsymbol{\gamma})}$}
\label{equality}
\begin{align}
    &\ket{\phi_p(\boldsymbol{\beta}, \boldsymbol{\gamma})} = \text{H}_+ \Big[\prod_{k=1}^p \Big( \exp(-i  \beta_k H_z) \exp(- i  \gamma_k H_I) \Big) \ket{0}^{\otimes n} \Big] \nonumber \\
    &= \text{H}_+ \text{H}_+^\dagger \Big[\prod_{k=1}^p \Big( \text{H}_+ \exp(-i  \beta_k H_z) \text{H}_+^\dagger \text{H}_+\times \nonumber \\
    &\hspace{4cm}\times\exp(- i  \gamma_k H_I) \text{H}_+^\dagger \Big) \text{H}_+ \ket{0}^{\otimes n} \Big] \nonumber\\
    &= \prod_{k=1}^p \Big(  \exp(-i  \beta_k \text{H}_+ H_z \text{H}_+^\dagger)  \exp(- i  \gamma_k \text{H}_+ H_I \text{H}_+^\dagger) \Big) \ket{+}^{\otimes n} \nonumber \\
    &= \prod_{k=1}^p \Big(  \exp(-i  \beta_k H_x)  \exp(- i  \gamma_k H) \Big) \ket{+}^{\otimes n} \nonumber \\
    &= \ket{\psi_p(\boldsymbol{\beta}, \boldsymbol{\gamma})}
\end{align}

\section{Details on SK Hamiltonian minimization in symmetric configuration}
\label{appen:sym_analysis}

We construct a basis $\mathcal{M} $ in $\mathcal{V}$ such that any easy instance, that is the ones with symmetric states in its ground space, can be minimized by at least one state $\ket{\chi} \in \mathcal{M}$. The set $\mathcal{M}$ comprises of 20 vectors divided into three classes. The first two classes contain states of the form:
\begin{equation}
    \ket{s_j} = \frac{1}{\sqrt{2}} (\ket{t_j} + X^{\otimes n} \ket{t_j}),
\end{equation}
 where $t_j$ are bit-strings, albeit with additional conditions. For the first class the condition is $\ket{t_j} = \mathcal{R} \ket{t_j}$. Since $n=6$ only 4 states belong to this class. For the second class the condition is $\ket{t_j} = \mathcal{R} X^{\otimes n} \ket{t_j}$. Again there are 4 states that belong to this class. For the third class the states are of the form:
\begin{equation}
    \ket{r_j} = \frac{1}{2}(\ket{t_j} + X^{\otimes n} \ket{t_j} + \mathcal{R} \ket{t_j} + \mathcal{R} X^{\otimes n} \ket{t_j}),
\end{equation}
provided that $\ket{t_j}$ do not obey the two conditions mentioned before. A total of 12 such states were enumerated. We numerically confirm that all the basis vectors can be prepared with our ansatz, that is, $\mathcal{M}\subset\mathcal{S}$.




\section{Numerical details}
The large calculations were performed on  Zhores supercomputer \cite{zacharov2019zhores}.
We required to classify weather each one of the $2^{15}$ 6-qubit SK instances could be minimized with ansatz \eqref{eq:QAOA_ansatz_exp}. 
In order to reduce the number of calculations, each time the code took a new instance its ground space was compared with those in a database. In the database the ground spaces were classified on whether they can be prepared by the ansatz or not. If the ground space was found in the database, the instance would be classified without requiring minimization.
On the contrary, if the ground space was not found in the database, the program would attempt to minimize the instance with up to 20 layers and its ground space would be added to the database.

The minimization process was performed using a layerwise inspired heuristic \cite{akshay2022circuit}. It starts with a single layer that is optimized, then keeping all the previous parameters fixed, a new layer is added and trained. This is followed by simultaneous optimization of all the parameters and the process of adding and training is repeated. In our implementation, at each optimization step we took the best of 25 seeds of the L-BFGS-B optimizer. The heuristic was repeated a total of 50 times in hopes of finding the global minimum.

\end{document}